\documentclass[aps,prb,twocolumn, superscriptaddress, showpacs]{revtex4}
\pdfoutput=1

\usepackage{graphicx, color}
\usepackage{dcolumn}
\usepackage{bm}
\usepackage{amsmath}
\usepackage{amssymb}
\usepackage{verbatim}
\usepackage{appendix}

\newcommand{\ket}[1]{|#1\rangle}               

\begin{document}


\title{Optical Polarization of $^{13}$C Nuclei in Diamond through Nitrogen-Vacancy Centers}


\author{Jonathan P. King}
\email[]{jpking@berkeley.edu}
\affiliation{Department of Chemical Engineering, University of California, Berkeley, California 94720, USA}
\author{Patrick J. Coles}
\affiliation{Department of Physics, Carnegie Mellon University, Pittsburgh, Pennsylvania 15213, USA}
\author{Jeffrey A. Reimer}
\affiliation{Department of Chemical Engineering, University of California, Berkeley, California 94720, USA}



\date{\today}

\begin{abstract}
We determine the polarization of the bulk $^{13}$C nuclear spin system in diamond produced by interaction with optically oriented nitrogen-vacancy (NV-) defect centers.  $^{13}$C nuclei are polarized into the higher energy Zeeman state with a bulk-average polarization up to 5.2\%, although local polarization may be higher.  The kinetics of polarization are temperature independent, and occur within 5 minutes.  Fluctuations in the dipolar field of the NV- center spin bath are identified as the mechanism by which nuclear spin transitions are induced near defect centers.  Polarization is then transported to the bulk material via spin diffusion, which accounts for the observed kinetics of polarization.  These results indicate control over the nuclear spin bath, a methodology to study dynamics of an NV- center ensemble, and application to sensitivity-enhanced NMR.
\end{abstract}


\maketitle

In recent years, much attention has been given to the negatively charged nitrogen-vacancy (NV-) defect in diamond.  This interest is due to its ground-state triplet, long coherence times, large spin polarization via optical pumping, and optical spin-state readout.  Most studies have focused on quantum control and the use of NV- centers as qubits for quantum information \cite{coupling,manipulation,coupledspins,dynamics,register,bath,entanglement}, while others have focused on applications such as high-resolution magnetometry\cite{Magnetometry1,Magnetometry2}.  A topic of particular interest is the interaction of NV- centers with the surrounding ``spin bath," which contains both nuclear and electronic spins\cite{bath1,bath2,bath3,quench}. 

We have investigated the NV- defect in diamond from the perspective of bulk $^{13}$C nuclear polarization in the presence of a large, static Zeeman field.  Previous work has shown microwave-induced dynamic nuclear polarization (DNP) of bulk $^{13}$C spins in diamond through paramagnetic defects\cite{DNP1,DNP2,DNP3}.  These DNP techniques are limited, however, by the electron Boltzmann factor at the magnetic field strength and temperature used in the experiment.  Recently, experiments have shown polarization of the nitrogen nuclear spin of the NV- center, as well as polarization of proximate single $^{13}$C spins by making use of the optical polarization of the NV- center\cite{singleDNP,register}.  These studies involved polarization and detection of single nuclear spins with optical methods.

In this Brief Report, we demonstrate an all-optical method of polarizing the bulk $^{13}$C (spin$\frac{1}{2}$) nuclear spin system in diamond at 9.4 Tesla and cryogenic temperatures.  We propose a mechanism for the polarization which relies upon the fluctuating dipolar field of the NV- center spin bath to induce nuclear spin transitions.  The equilibrium behavior of the polarization is modeled by invoking the spin-temperature concept, where the nuclear spin temperature equilibrates with the temperature of the dipolar reservoir of the NV- center spin bath.

The sample used in this study is a 3x3x0.5mm high pressure, high temperature (HPHT) synthetic single-crystal diamond purchased from Element-6.  After irradiation treatment and annealing, the final concentration of NV- centers was 8 ppm ($\sim1.4*10^{18}$cm$^{-3}$).\footnote{This sample preparation is detailed in Ref. 18, where it is labeled sample 6.}  

NMR experiments were carried out in a 9.4 Tesla superconducting magnet, where spectra were acquired with a single-resonance spectrometer at 100.591 MHz.  The sample was mounted in a homebuilt NMR probe employing a flattened split-solenoid coil for optical access.  A $\langle 111 \rangle$ crystal axis was aligned with the magnetic field.   Temperature control was maintained using an Oxford Spectrostat continuous flow helium cryostat with an Oxford ITC-503S temperature controller.  Optical pumping was performed using an argon ion laser operating simultaneously at multiple wavelengths ranging from 457.9nm to 514.5nm.  Power output from the laser was 2.5 Watts except where laser-power dependence was studied.   Optical pumping experiments were performed by saturation of the $^{13}$C spin transitions via a series of 50 $\frac{\pi}{2}$ pulses, each followed by a 10 ms dephasing time.  After saturation, a variable period of optical pumping occurs, then a $\frac{\pi}{2}$ pulse immediately followed by detection of the free-induction decay (FID).  Laser irradiation was maintained throughout the entire experiment.  Non optically-pumped thermal equilibrium spectra were acquired by cooling the sample to the specified lattice temperature, allowing the spins to equilibrate for two hours (much longer than the nuclear $T_1$), and applying a $\frac{\pi}{2}$ pulse to detect the FID.  The integrated area of the Fourier transformed FID is taken to be proportional to the bulk-average nuclear polarization.  Error from random noise was estimated from the root-mean-square magnitude and correlation time of the noise in the spectrum, and was found to be negligible.  The error that is apparent in the data with less signal is likely due to imperfect baseline and phase corrections.  Lorentzian curves accurately fit the lineshapes and were used to obtain peak shifts and widths.

\begin{figure}
\vspace*{-1.0cm}
\includegraphics[width=0.5\textwidth]{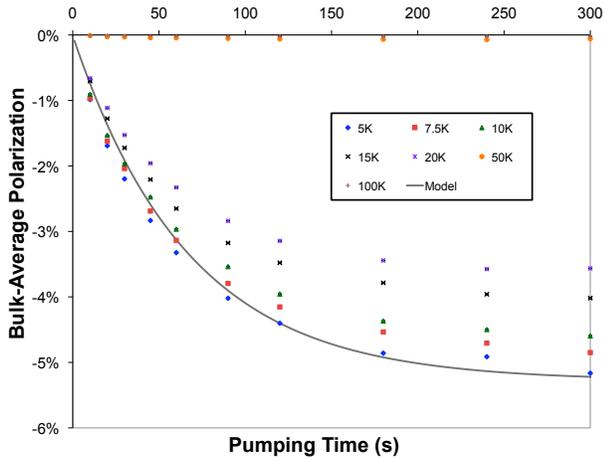}
\caption{\label{temp}(Color online) $^{13}$C polarization kinetics with 2.5W irradiation and varied temperature. The theoretical model is fit to the 5K data.}
\end{figure}

\begin{figure}[htp]
\vspace*{-1.0cm}
 \centering
    \includegraphics[width=.5\textwidth]{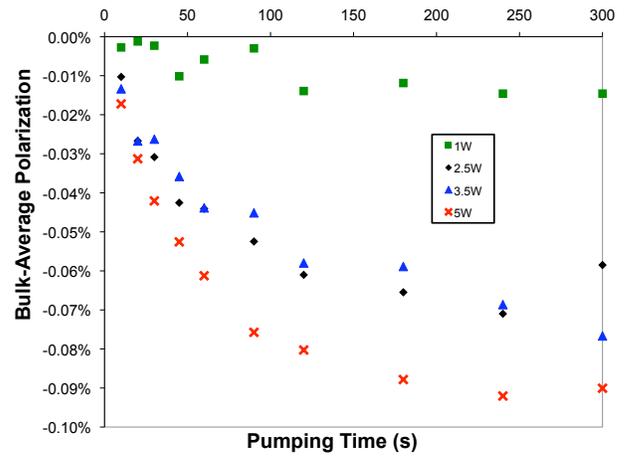}
     \caption{(Color online) Polarization kinetics as a function of laser power at 50K.}
     \label{power}
\end{figure}

In all experiments, optical pumping resulted in $^{13}$C NMR signals opposite in sign to thermal equilibrium.  The magnitude of the polarization increased with lower temperature (Fig. \ref{temp}) and larger irradiation power (Fig. \ref{power}), with a maximum bulk-average polarization of 5.2\%.  As discussed later, the illuminated volume is less than the total volume of the sample, so the local polarization is significantly greater than 5\%.  The polarization rate appears to be identical in all cases.


The NMR peak width and frequency shift showed a modest temperature dependence, with derived values given in Table \ref{shift}.  It should be noted that significant laser heating of the sample may occur, and for temperatures 20K and lower, the stability of temperature control was poor due to the limited cooling capacity of the cryostat.

%

\begin{table}[htp]
\begin{center}
  \begin{tabular}{| l | c | r | }
    \hline
    Temperature & Peak Width & Peak Shift \\ \hline
   5K & 776Hz & -363Hz \\ \hline
   7.5K & 757Hz & -315Hz \\ \hline
  10K & 762Hz & -272Hz \\ \hline
   15K & 731Hz & -116Hz \\ \hline
   20K & 698Hz & 0Hz \\ 
    \hline
  \end{tabular}
\end{center}
\caption{NMR peak width and shift relative to the peak at 20K, obtained by fitting Lorentzian curves to the data.}
\label{shift}
\end{table}

These experimental results give insight into the mechanism of the polarization process.  The lack of temperature dependence of the polarization rate, along with the presence of a large Zeeman field and low temperature, suggests a process which does not require energy exchange with the crystal lattice.  We therefore consider energy-conserving transitions between $^{13}$C Zeeman and NV- center dipolar energy levels.  Similar processes have been examined as spin-lattice relaxation mechanisms in diamond and other materials containing paramagnetic centers\cite{relaxation,poulis}.


The total spin Hamiltonian for the system may be written as:
\begin{equation}
\mathcal{H}=\mathcal{H_S}+\mathcal{H_{IS}}+\mathcal{H_I}\\
\end{equation}

\noindent$\mathcal{H_S}$ refers to terms involving only NV- centers, $\mathcal{H_I}$ involves only nuclear spins, and $\mathcal{H_{IS}}$ represents the interactions between the two.  In this treatment, $\mathcal{H_{IS}}$ is a small perturbation that couples the two energy reservoirs independently defined by $\mathcal{H_S}$ and $\mathcal{H_I}$.  We consider a simple model that system $S$ is composed of two NV- defects.  Including pairwise dipolar interactions (between defects labeled 1 and 2 ) and retaining only those terms that commute with the dominant Zeeman Hamiltonian:
\begin{eqnarray}
\mathcal{H_S}=\hbar\omega_S(S_{1,Z}+S_{2,Z})+\Gamma_1(S_{1,Z})^2+\Gamma_2(S_{2,Z})^2\\
+AS_{1,Z}S_{2,Z}-\frac{1}{4}A(S_{1,+}S_{2,-}+S_{1,-}S_{2,+}) \nonumber
\end{eqnarray}
where $A=\frac{\mu_0}{4\pi}\hbar^2\gamma^2_{NV}\frac{1}{r^3}[1-3\cos^2\theta]$ quantifies the electron-electron dipolar interaction between defects, and $\Gamma_i=2.88\mbox{GHz}*2\pi\hbar*\frac{1}{2} \left(3\cos^2\gamma_i-1 \right)$ is the magnitude of the spin-spin interaction within a single defect\cite{loubser}, which we have indexed to account for different defect orientations.  $\theta$ and $r$ specify the orientation and length of the vector connecting the two NV- centers and $\gamma$ is the angle of a defect symmetry axis with the external magnetic field (0 or 109.5 degrees due to the tetrahedral symmetry of the crystal).  Diagonalization of $H_S$ gives the energy levels for the NV- centers.  In order for the perturbation $\mathcal{H_{IS}}$ to induce nuclear spin transitions, an accompanying transition at the nuclear Zeeman energy must take place within the energy levels of $\mathcal{H_S}$. Due to the large Zeeman energy ($\sim$263 GHz) of NV- centers, transitions between NV- center Zeeman levels can never be ``on resonance" with a nuclear spin transition ($\sim$100MHz).  Therefore, only transitions for which $\Delta S_Z=0$ may participate.  Considering the 9 Zeeman eigenstates, the $S_Z=\pm2$ states are energetically isolated, the $S_Z=\pm1$ manifolds induce nuclear transitions but are unpolarized, and the $S_Z$=0 manifold induces nuclear transitions and is polarized.  The energy eigenstates with $S_Z$=0 have energies:
\begin{eqnarray}
E_1&=&-A+\Gamma_1+\Gamma_2 \\
E_2&=&\frac{1}{2}(-A-\sqrt{2A^2+(A-\Gamma_1-\Gamma_2)^2}+\Gamma_1+\Gamma_2)\\
E_3&=&\frac{1}{2}(-A+\sqrt{2A^2+(A-\Gamma_1-\Gamma_2)^2}+\Gamma_1+\Gamma_2)
\end{eqnarray}
Since $(\Gamma_1+\Gamma_2)$ is large compared to both the nuclear Zeeman energy,  $\hbar\omega_I$, and the magnitude of the dipolar interaction, $A$, only transitions between $E_1$ and $E_3$ have an appreciable probability of yielding the necessary energy for the 100 MHz nuclear transition.
In the two-defect Zeeman basis, the state vectors for these states are:
\begin{eqnarray}
\ket{1}&=&\frac{1}{\sqrt{2}}(-\ket{-1,1}+\ket{1,-1})\\
\ket{3}&=&\frac{1}{N_3}(\ket{-1,1}+\alpha\ket{0,0}+\ket{1,-1})
\end{eqnarray}
\noindent where $N_3=\sqrt{2+\alpha^2}$ and $\alpha=-2E_2/A$.  Since the $\ket{0,0}$ Zeeman state is known to be selectively populated by optical pumping,\cite{measurement} the energy eigenstate $\ket{3}$ will become selectively populated.  The energy eigenstates $\ket{1}$ and $\ket{3}$ are taken to be a pseudo-two level system, and are assigned a temperature:
\begin{equation}
T=\frac{E_3-E_1}{k_B\ln(\frac{P_1}{P_3})}
\label{spintemp}
\end{equation}
For defect pairs oriented such that $E_3>E_1$, selective population of state $\ket{3}$ will result in a negative temperature, and thermal contact of this pseudo-two level system with the nuclear spins drives the nuclear spins to a negative temperature.  After calculating transition rates, it is apparent these negative temperature transitions dominate, accounting for the observed NMR signal.





The preceding model may be further quantified by considering the rates of transitions between the pseudo-two level system formed by  
pairs of NV- centers and the $^{13}$C nuclear spins.  Towards that  
end, we first must identify the relationship between the known  
populations of the three Zeeman eigenstates ($P_{\ket{+1}}, P_{\ket{0}},  
P_{\ket{-1}}$)  formed by the {\it single} NV- center, and the ratio of  
the two-defect populations, $P_1/P_3$.  Consideration of  
the 9x9 density matrix for  
coupled NV- centers yields
\begin{eqnarray}
P_{1}&=&P_{\ket{+1}}P_{\ket{-1}}\\
P_{3}&=&\frac{1}{N_3^2}\left(2P_{\ket{+1}}P_{\ket{-1}}+\alpha^2P^2_{\ket{0}}\right)
\end{eqnarray}


The dipole-dipole interaction between NV- centers and $^{13}$C spins is treated as a perturbation that induces transitions between NV- center energy levels.  Fermi's Golden Rule gives the direct relaxation rate for a nucleus, \textit{m},  near a central NV- center, \textit{j}, which is also coupled to another NV- center, \textit{k}:
\begin{equation}
W^{jkm}=\frac{2\pi}{\hbar}|\mathcal{V}^{jm}|^2\delta(\Delta E-\hbar\omega_I)
\end{equation}
$\mathcal{V}^{jm}=\frac{\mu_0}{4\pi}\hbar^2\gamma_{NV}\gamma_{I}\frac{1}{r^3}(-\frac{3}{2}sin(\theta)cos(\theta)e^{-i\phi})$ is the Hamiltonian matrix element for the transition, from $\mathcal{H_{IS}}$.  Angular dependence is removed by spatial averaging\footnote{This was done in Ref. 19 and is valid when spin diffusion, not direct polarization, is the dominant mode of transport.} to obtain $|\mathcal{V}^{jm}|=-\frac{\mu_0}{4\pi^2}\hbar^2\gamma_{NV}\gamma_{I}\frac{1}{r^3}$.  In order to obtain an average transition rate, a sum over all neighboring NV- centers, \textit{k}, is approximated as an integral:
\begin{equation}
W^{jm}=\int_{-\infty}^\infty \rho(A) g(A)W^{jmk}(A) \, dA
\end{equation}
where $g$ is the normalized distribution of values of A, and $\rho$ is the probability that the NV- center pair is in one of the states participating in the transition.  For a dilute spin system, $g(A)=\frac{\delta}{\pi(\delta^2+A^2)}$, where $\delta=\frac{2\pi^2}{3\sqrt{3}}\gamma_{NV}^2\hbar^2 n$ and $n$ is the number density of NV- centers \cite{abragam}.  The integral may then be evaluated:
\begin{equation}
W^{jm}=\sum_{n=1,2}\rho(A_n)W^{jmk}(A_n)g(A_n)
\end{equation}
\noindent where $A_n$ are the two roots of the argument of the delta function in $W$.  

In the present model, direct relaxation of $^{13}$C nuclei occurs only near NV- centers.  Thus bulk $^{13}$C signals build up due to transport of the polarization via nuclear spin diffusion.  The governing equation for such spin transport is:
\begin{equation}
\frac{\partial P}{\partial t}=D\frac{1}{r^2}\frac{\partial}{\partial r}(r^2\frac{\partial P}{\partial r})+\sum_{l}f_lW_l(P_l-P)-W_0P
\end{equation}
\noindent where $D=6.7*10^{-15}cm^2/s$ is the spin diffusion coefficient due to the homonuclear dipolar interaction $\mathcal{H_{II}}$\cite{relaxation}, neglecting hyperfine ``shifts" near the defect\cite{dynamics}.  $P$ is the nuclear spin polarization, $P_l$ is the nuclear polarization in equilibrium with the temperature in Eq. (8), $W_l$ is the direct relaxation rate and $f_l$ is the fraction of NV- center pairs in a given orientation, $l$.  $W_0$ is the total transition rate within the unpolarized $S_Z=\pm1$ manifolds.  Nuclear spin-lattice relaxation is neglected in the calculations.  Using the initial condition $P(t=0)=0$ for saturation of the spin transitions, and the boundary conditions $\frac{dP}{dr}(r=0)=0$ and $\frac{dP}{dr}(r=R)=0$, which follow from symmetry, a numerical solution may be obtained.  Using $n=9*10^{18}$cm$^{-3}$ , the numerical solution was rescaled and superimposed upon data taken at 5K in Fig. \ref{temp}.  This solution requires a single-defect NV- center polarization of more than $99\%$ so that the $S_Z=\pm1$ transitions do not dominate depolarization of the nuclei.  This polarization is higher than the $80\%$ reported in literature\cite{measurement}.  The present sample, however, has a larger NV- concentration and our measurements were made in a larger magnetic field, such that the polarization may be higher than reported in Ref. 21.

The time scales apparent in Figure 1 may be compared to estimates assuming Eqn. 14  is governed exclusively by the
diffusion coefficient D. The 8ppm defect concentration presented in Ref. 18 may be used to predict the characteristic diffusion time, 
$\tau=R^2/D$.  The resulting time-scale, $\sim45$ seconds, is of the same order of magnitude as the data shown in Fig. 1. Thus, in our model, the contribution of NV- center dipolar fluctuations (the rate processes given in Eqn. 13) to the overall process reflect a delicate balance of spin-diffusion, direct relaxation rates, and transitions within the unpolarized Sz=+/- 1 levels.   

The equilibrium behavior as described by the spin-temperature treatment is consistent with the anti-thermal sign of the NMR signal.  The polarization kinetics are consistent with direct relaxation near defects with spin diffusion transport of polarization to the bulk material.  The origin of the temperature dependence of the ultimate polarization is not as clear.  For many nuclear polarization methods, temperature dependance is due to either the Boltzmann factor or ``leakage" of polarization via nuclear spin-lattice relaxation.  The lack of temperature dependence of the polarization kinetics is not consistent with this picture.  Instead, we propose the temperature dependence is due to a decreased laser penetration depth at higher temperatures.  Since irradiation is occurring within the phonon sideband of the NV- center optical absorption spectrum\cite{acosta}, an increase in absorption with increasing temperature is expected as more phonons are available to participate in the absorption, leading to a decrease in optical penetration and a smaller irradiation volume.  Thus, a model for the temperature and power dependence, while beyond the scope of this article, could consider their effect on the spatial profile of light intensity $I(z)$, and in turn, the polarization.


The present model, as summarized in Equation 14, does not account for all contributions to the spectral density of fluctuating 
fields at the $^{13}$C nucleus. The temperature-dependence of the $^{13}$C lineshapes suggests that the are other factors influencing the spin dynamics in this sample, likely other paramagnetic defects such as substitutional nitrogen, which have highly temperature dependent polarization and are also coupled to the nuclei and may contribute to relaxation processes. This and other effects, for example, the dynamics of NV- center photophysics, would effectively result in a modification of the function g(A).  However, our simple model accounts for the dominant features of the data, and may be further tested by noting that the temperature in Eq.\eqref{spintemp} depends on the crystal orientation through the parameter $\Gamma$.

We consider the results of this study to be important to several scientific communities.  The mechanism described herein portends control over the nuclear spin bath in diamond, whose interaction with NV- centers is an area of active research.  This work also suggests that bulk $^{13}$C NMR measurements probe the dynamics of an ensemble of NV- centers, providing a methodology complimentary to optical (single NV- center) techniques.  Spin-polarized nuclei in diamond may also be useful for sensitivity-enhanced NMR techniques where polarization localized near the crystal surface may be transfered to other nuclei (e.g. adsorbed molecules) for enhanced NMR sensitivity.

We thank Dmitry Budker for providing the sample used in this study.  We also wish to thank Rachel Segalman for support and Carlos Meriles for helpful discussion.  J.P.K. acknowledges funding from the NDSEG Fellowship.  The authors acknowledge support from the National Science Foundation  
under project
ECS-0608763.

\end{document}